\documentclass[aps]{revtex4-2}
\usepackage{graphicx}
\begin{document}

\preprint{}

\title{Stochastic resonance in periodically driven bistable systems subjected to anomalous diffusion} 



\author{F. Naha Nzoupe}%

\affiliation{Laboratory of Research on Advanced Materials and Nonlinear Science (LaRAMaNS), Department of Physics, Faculty of Science, University of Buea P.O. Box 63 Buea, Cameroon.}

\author{Alain M. Dikand\'e}
\email[Corresponding author: ]{dikande.alain@ubuea.cm}
\affiliation{Laboratory of Research on Advanced Materials and Nonlinear Science (LaRAMaNS), Department of Physics, Faculty of Science, University of Buea P.O. Box 63 Buea, Cameroon.}

\date{\today}

\begin{abstract}
The occurrence of stochastic resonance in bistable systems undergoing anomalous diffusions, which arise from density-dependent fluctuations, is investigated with emphasis on the analytical formulation of the problem as well as a possible analytical derivation of key quantifiers of stochastic resonance. The nonlinear Fokker-Planck equation describing the system dynamics, together with the corresponding Ito-Langevin equation, are formulated. In the linear-response regime analytical expressions of the spectral amplification, of the signal-to-noise ratio and of the hysteresis loop area are derived as quantifiers of stochastic resonance. These quantifiers are found to be strongly dependent on the parameters controlling the type of diffusion, in particular the peak characterizing the signal-to-noise ratio occurs only in close ranges of parameters. Results introduce the relevant information that taking into consideration the interactions of anomalous diffusive systems with a periodic signal, can provide a better understanding of the physics of stochastic resonance in bistable systems driven by periodic forces. 
\end{abstract}

\maketitle 

\section{Introduction}
\label{intro}
Nonlinear systems exhibit a broad range of unexpected behaviors as well as complex dynamical properties, due to the presence of noise \cite{1,2,3,3a,risk}. As typical example, thermally activated processes such as noise-induced escape from metastable states \cite{1,2,3,3a,risk}, and the phenomenon of stochastic resonance (SR) \cite{3,4,nic}, have attracted a great deal of attention because of their fundamental role in several areas of physics, chemistry, biophysics, social as well as financial sciences \cite{risk,kim,berg,gei}. In particular there has been an unprecedented interest in noise-driven phenomena pertaining to the so-called Kramers escape rate problem \cite{risk}, due to their intimate connection with second-order phase transitions \cite{schrif} in physical contexts where dynamical instabilities are governed by double-well potentials \cite{dik,dika,dik1}. For these systems noise-driven transport phenomena are well-known to be dominated by standard diffusion processes, whereby oscillators exhibit normal Brownian motions and the escape rate has the standard Arrhenius law \cite{3}. Investigations of SR in systems of this class have led to the important conclusion that noise multiplicativity and time correlation are necessary for the occurrence of SR in linear systems, and that nonlinearity is an essential ingredient for the occurrence of SR in physical contexts with additive white noise \cite{berd,berd1,berd2,berd3,berd3a}. Also required is the interplay of at least two possible equilibrium states with sufficiently confining strength and separated by a potential barrier \cite{11,12,13}. \\
In nature we equally encounter physical systems in which noise-driven transports involve diffusion phenomena that deviate strongly from the classic picture. Indeed certain diffusion phenomena exhibit nonlinear transport features which are completely distinct from the standard Brownian motion, resulting in an anomalous behavior of the escape dynamics \cite{sok1,sok2,sok3,yina,nep}. Moreover they display different scenarios of occurrence of SR, as evidenced by some recent results on this issue \cite{sok1,sok2,sok3}. \\
In this work we are interested in SR for systems exhibiting correlated-type anomalous diffusions, in particular we focus on physical contexts where diffusion processes depend on the probability density $P$ of particles. Here we treat the case where the diffusion coefficient is proportional to an arbitrary power of $P$, such that depending on values of two power coefficients $\nu$ and $\mu$ the system is either subdiffusive \cite{yina,subdif1} or superdiffusive \cite{yina,subdif1}. Ranging from cell dynamics \cite{difa}, surface growth \cite{14}, radiative heat transfert by Marshak waves \cite{15}, gravitational spreading of thin liquid films \cite{16}, spatial diffusion of biological populations \cite{17} and saturation of thin regions in porous media \cite{18}, there exists a wealth of real physical situations were stochastic processes display anomalous diffusions with a density-dependent diffusion coefficient. Some expressions for the escape time were proposed in the literature \cite{19,20}, and were shown to be strongly dependent on the order of anomaly of the system diffusion. To our knowledge no theoretical investigation of the occurrence of SR in such systems has been performed. It would therefore be particularly interesting to know how systems  exhibiting density-dependent diffusion coefficients would respond to an external periodic forcing. First of all we shall discuss the Kramers escape problem for bistable systems with anomalous diffusion, ignoring the periodic forcing. then we examine the effects of a sinusoidal-type periodic driving force on the occurrence of SR in the system.   
\section{Model and escape rate in the presence of anomalous diffusion}
\label{sectwo}
The class of systems we are interested in can be represented as set of particles immersed in a thermal bath. In order to probe the escape behavior, and the possible occurrence of SR in the systems, we introduce a bistable potential coupled to a periodic forcing. At the macroscopic scale the system dynamics can be described by the following nonlinear Fokker-Planck equation \cite{yina,nep}:
\begin{equation}
\partial_{t}P^{\mu}(x,t) = - \partial_{x}\left[f(x,t)P^{\mu}(x,t)\right]+ D\partial_{x}^{2}P^{\nu}(x,t),
\label{e1}
\end{equation}
where $f(x,t)= -\partial_{x}U(x) + A(t)$ is the external force with $U$ a bistable potential of the form $U(x)= U_{0}(x^{2}-1)^{2}$, and $A(t)$ a sinusoidal periodic forcing i.e. $A(t)= A_{0}sin\omega t$. $U_0$ and $A_0$ are the potential barrier and the magnitude of the periodic force, respectively. The paramater $D$ in eq. (\ref{e1}) measures the anomalous diffusion-induced noise strength, while $(\mu,\nu)$ are real and positive parameters characterizing the deviation of the diffusion process from the normal diffusion. \\ Eq. (\ref{e1}) describes a driven bistable system undergoing diffusion processes, which can be either a superdiffusion when $\mu > \nu$, a normal diffusion when $\mu = \nu$ or a subdiffusion when $\mu < \nu$ \cite{sok3,yina}. Instructively the double-well potential $U(x)$ has its stable states located at $x_{\pm}=\pm 1$, with a barrier $E_{b}= U_{0}$ centered at $x_{b}=0$. We first examine the Kramers escape problem in the absence of periodic driving. When $A_0=0$, the system dynamics is mainly governed by random fluctuations and may lead to a transition between the states $x_{\pm}=\pm 1$. The Fokker-Planck equation (\ref{e1}) in this case reduces to:
\begin{equation}
\partial_{t}P^{\mu}(x,t) = \partial_{x}\left[P^{\mu}(x,t)\frac{dU(x)}{dx}\right]+ D\partial_{x}^{2}P^{\nu}(x,t).
\label{e2}
\end{equation}
By optimizing the entropy of the system \cite{tsallis}, the stationary solution to eq. (\ref{e2}) is straightforward yielding:
\begin{equation}
P_{st}(x)= \frac{1}{Z}\left[ 1 - \left(\frac{\nu-\mu}{\mu}\right)\beta_{\mu\nu}U(x)\right]^\frac{\mu}{\nu -\mu}_{+}, \label{e3}
\end{equation}
where $\left[.\right]_+= max \lbrace ., 0\rbrace$, $\beta_{\mu\nu}$ defined by:
\begin{equation}
\beta_{\mu\nu} = \frac{\mu}{\nu D}Z^\frac{\nu-\mu}{\mu}, 
\label{e4}
\end{equation} 
and the normalization constant $Z= \int \left[ 1 - \mu^{-1}\left(\nu-\mu\right)\beta_{\mu\nu}U(x)\right]^\frac{\mu}{\nu -\mu}_{+}.$
For a nonlinear medium with arbitrary diffusion characterized by values of $(\mu,\nu)$, it is useful \cite{19,vil} to define a suitable $(\mu,\nu)$-dependent effective potential, which in the present context reads:
\begin{equation}
\phi_{e}(x)= \frac{\nu}{\mu -\nu}ln\left[ 1 - \left(\frac{\nu-\mu}{\mu}\right)\beta_{\mu\nu}U(x)\right].
\label{e5}
\end{equation}
In terms of this effective potential, the exact expression for the current of particles across the effective-potential barrier, i.e. $S$, will be: 
\begin{eqnarray}
S(x,t) = -D\exp\left[-\frac{\phi_{e}(x)}{D}\right]\frac{\partial}{\partial x}\left[\exp\left(\frac{\phi_{e}(x)}{D}P^{\nu/\mu}(x,t)\right)\right]. \nonumber \\
\label{e6}
\end{eqnarray}
By invoking Kramers rate theory \cite{3} we obtain the escape rate in the small-noise limit i.e.:
\begin{eqnarray}
r_{k} &\approx& \frac{Z^\frac{\nu-\mu}{\mu}\sqrt{U''(x_{\pm})| U''(x_{b})|}}{2\pi}\nonumber \\\
&\times&\left[ 1-\left(\frac{\nu-\mu}{\mu}\right)\beta_{\mu\nu} U(x_{b})\right]_{+}^{\frac{\mu+\nu}{2(\nu-\mu)}}\nonumber \\
&\times&\left[ 1-\left(\frac{\nu-\mu}{\mu}\right)\beta_{\mu\nu} U(x_{\pm})\right]_{+}^{\frac{\mu-3\nu}{2(\nu-\mu)}}, \nonumber \\
\label{e7}
\end{eqnarray} 
where $U''= d^{2}U/dx^{2}$. For the double-well potential $U(x)$ defined above, eq. (\ref{e7}) simplifies to:
\begin{equation}
r_{k} \approx \frac{2\sqrt{2}E_{b}Z^\frac{\nu-\mu}{\mu}}{\pi}\left[ 1-\left(\frac{\nu-\mu}{\mu}\right)\beta_{\mu\nu}E_{b}\right]_{+}^{\frac{\mu+\nu}{2(\nu-\mu)}}.
\label{e8}
\end{equation}
This is the Kramers escape rate for a bistable system subjected to anomalous diffusion. \\
Let us now examine the system dynamics taking into consideration the periodic forcing. At the microscopic scale, where the distributions of particle positions in time is meaningful, we need to consider the Ito-Langevin equation \cite{ito1,ito2,ito3} coupled to eq.(\ref{e1}), which in the present specific context is expressed:
\begin{equation}
\frac{\partial x }{\partial t} = -f(x,t) + g(x,t)\eta(t),
\label{e9}  
\end{equation}
with $\eta(t)$ standing for for thermal fluctuations. From eq.(\ref{e1}) the probability distribution of a particle position $x$ at time $t+\Delta t$ must satisfy \cite{risk}:
\begin{equation}
P^{\mu}(x,t+\Delta t) = \int p(x,t+\Delta t|x',t)P^{\mu}(x',t)dx',
\label{e10}  
\end{equation}
where $p$ denotes the transition probability. Setting $x=x' + \Delta x$, a Taylor expansion of the integrand in eq. (\ref{e10}) for small $\Delta x$ leads to the relation:
\begin{equation}
\langle\Delta x^{n}\rangle = \int \Delta x^{n}p(x+\Delta x,t+\Delta t|x,t)d\Delta x,
\label{e11}  
\end{equation}
from which the Kramers-Moyal expansion coefficient can be derived \cite{21}. Moreover based on eq. (\ref{e11}) it was established that to have a correspondence between eq. (\ref{e1}) and eq. (\ref{e9}), with $\eta(t)$ obeying $\langle\eta(t)\rangle = 0$ and $\langle\eta(t)\rangle\langle\eta(s)\rangle = \delta(t-s)$, the following relation should hold: 
\begin{equation}
\frac{1}{2}g^{2}(x,t) = D\left[P(x,t)\right]^{\nu-\mu}.
\label{e12}  
\end{equation}
Using this constraint, we can rewrite the Ito-Langevin equation describing the system dynamics at the microscopic scale as:
\begin{equation}
  \frac{d x }{dt} = -\frac{\partial U(x)}{\partial x} + A_{0}sin\omega t + \sqrt{2D}\left[P(x,t)\right]^{\frac{\nu-\mu}{2}}\eta(t).
\label{e13}  
\end{equation}
In the case of normal diffusion (i.e. $\mu = \nu$) eq. (\ref{e14}) reduces to the Langevin equation for a constant noise. The state dependence of noise in eq. (\ref{e14}) can be interpreted as a consequence of interactions with environment that cannot be explicitly taken into account by the equations of motion. Equation (\ref{e14}) thus provides a phenomenological description in which the macroscopic quantity $P$, stands for a kind of statistical feedback determining microscopic trajectories for the case of anomalous diffusion $(\mu \neq \nu)$. During this process the interactions with the environment are such that as the system evolves, the particle is affected  by the surrounding collective density of states.
\section{Stochastic resonance in the presence of periodic driving}
\label{three}
We set the system in the small-noise regime and weak-modulation strength, i.e. $ \vert A_{0}x_{s}\vert \ll E_{b}$ and $\omega \ll U''(x_{\pm})$, such that there are no possible transitions between stable states in the absence of noise. Ignoring intrawell dynamics in the Ito-Langevin equation (\ref{e13}), the system response in the long-time regime is governed by its harmonic component i.e. \cite{3}:
\begin{equation}
\langle x(t) \rangle = \bar{x}(D)\sin(\omega t - \bar{\phi}(D)),
\label{e14}  
\end{equation}
where the amplitude $\bar{x}$ and phase lag $\bar{\phi}$, in the linear-response approximation, are given by:
\begin{equation}
\bar{x}(D) = \left( \frac{ 3\nu - \mu }{2\mu}\right)\frac{2A_{0}\beta_{\mu\nu}\langle x^{2} \rangle_{0}r_{k}}{\left(4r_{k}^{2} + \omega^{2} \right)^{1/2}},
\label{e15}  
\end{equation}
and
\begin{equation}
\bar{\phi}(D) = \arctan\left( \frac{\omega}{2r_{k}}\right),
\label{e16}  
\end{equation}
with $\langle x^{2} \rangle_{0}$ the variance of the stationary unperturbed process. The spectral power amplification (SPA) \cite{herm1,herm2}, defined as the ratio of the power of the driven oscillation to the power of the driving signal at the driving frequency $\omega$, is obtained as: 
\begin{eqnarray}
SPA &=& \left[\frac{\bar{x}(D)}{A_{0}} \right]^{2} \nonumber \\
&=& \left[\left( \frac{\mu - 3\nu}{2\mu}\right)\beta_{\mu\nu}\langle x^{2} \rangle_{0}\right]^{2}\frac{4r_{k}^{2}}{\left(4r_{k}^{2} + \omega^{2} \right)}.
\label{e17}  
\end{eqnarray} 
\\
Using the theory of SR \cite{3} and looking at the problem in terms of output signal power spectrum, the phase-averaged power spectral density is given by \cite{3,herm1}:
\begin{eqnarray}
S(\Omega) &=& (\pi/2)\bar{x}(D)^{2}\left[\delta(\Omega+\omega) + \delta(\Omega-\omega) \right] \nonumber \\
&+& \left[ 1-\left( \frac{\mu - 3\nu}{2\mu}\right)^{2}\frac{4A_{0}^{2}\beta_{\mu\nu}^{2}\langle x^{2} \rangle_{0}r_{k}^{2}}{4r_{k}^{2} + \omega^{2}} \right]\frac{4r_{k}^{2}\langle x^{2} \rangle_{0}}{4r_{k}^{2} + \omega^{2}}. \nonumber \\
\label{e18}  
\end{eqnarray}
If we neglect terms of orders higher than $(A_{0})^{2}$, the leading term in the expression of the signal-to-noise ratio (SNR) \cite{3} thus reads:
\begin{equation}
SNR = \pi\left[\left( \frac{\mu - 3\nu}{2\mu}\right)A_{0}\beta_{\mu\nu}\langle x^{2} \rangle_{0}\right]^{2}r_{k} + \mathcal{O} ((A_{0})^{3}).
\label{e19}  
\end{equation}
For a normal Brownian motion for which $\mu=\nu$, eqs. (\ref{e15}), (\ref{e17}) and (\ref{e19}) clearly meet the expressions proposed in the literature \cite{3}. Moreover from conservation law the work done over a period $\tau = 2\pi/\omega$ of the periodic signal, equals the change in internal energy $\Delta E = f(x,\tau)-f(x,0)$ plus the heat $Q$ absorbed over a period, i.e. $W = \Delta E + Q$. Following Sekimoto's stochastic energetic formalism \cite{22}, the work done on the system over a period of the periodic forcing is defined as follows:
\begin{equation}
 W  = - \int_{0}^{\tau}x(t)\frac{dA(t)}{dt}dt. 
 \label{e20}
\end{equation}
The position $x(t)$ being a stochastic variable it is more accurate to express the work done in terms of its average value. Hence the average work done over a period is:
\begin{equation}
\langle W \rangle = - \int_{0}^{\tau}\langle x(t)\rangle\frac{dA(t)}{dt}dt. 
 \label{e21}
\end{equation}
With the expressions of $\langle x(t)\rangle$ given by eq. (\ref{e14}), $\bar{x}(D)$ given by eq. (\ref{e15}) and $\bar{\phi}(D)$ given by eq. (\ref{e16}), we find:
\begin{equation}
\langle W \rangle = \left( \frac{ 3\nu - \mu }{2\mu}\right)\frac{2\pi A_{0}^{2}\beta_{\mu\nu}\langle x^{2} \rangle_{0}\omega r_{k}}{4r_{k}^{2} + \omega^{2}}.
 \label{e22}
\end{equation}
The quantity $\langle W\rangle$ given by eq. (\ref{e22}) also represents the hysteresis loop area (HLA) of the system \cite{his} over a period of the forcing, while eqs. (\ref{e17}), (\ref{e19}) and (\ref{e22}) predict SR as a function of $D$ with a high dependence on the parameters $\mu$ and $\nu$ characterizing the diffusion anomaly. \\
Since we are dealing with a two-state system, the position distribution of the particle is the sum of two delta functions. The variance of position distributions having two delta functions equidistant from the origin is $\langle x^{2} \rangle_{0} = x^{2}_{\pm}=1$, for the bistable potential $U(x)$. With this we find that SR is expected at a noise strength:
 \begin{equation}
 D_{sr} = \left( \frac{5\nu-3\mu}{4\nu}\right)Z^\frac{\nu-\mu}{\mu}E_{b}.
  \label{e23}
 \end{equation}
The small-noise limit considered, i.e. $0 < D_{sr} \ll E_{b} $, yields the following condition from the noise strength $D_{sr}$ obtained in eq. (\ref{e23}):
 \begin{equation}
\frac{3}{5}\mu  < \nu < \frac{3Z^\frac{\nu-\mu}{\mu}}{5Z^\frac{\nu-\mu}{\mu} -4}\mu.
  \label{e24}
 \end{equation}
 Setting $Z=1$, the later condition turns to $(3/5)\mu <\nu < 3\mu$.
 
\section{Results and discussions}
\begin{figure}\centering
\begin{minipage}{0.5\textwidth}
\includegraphics[width=2.5in,height=2.2in]{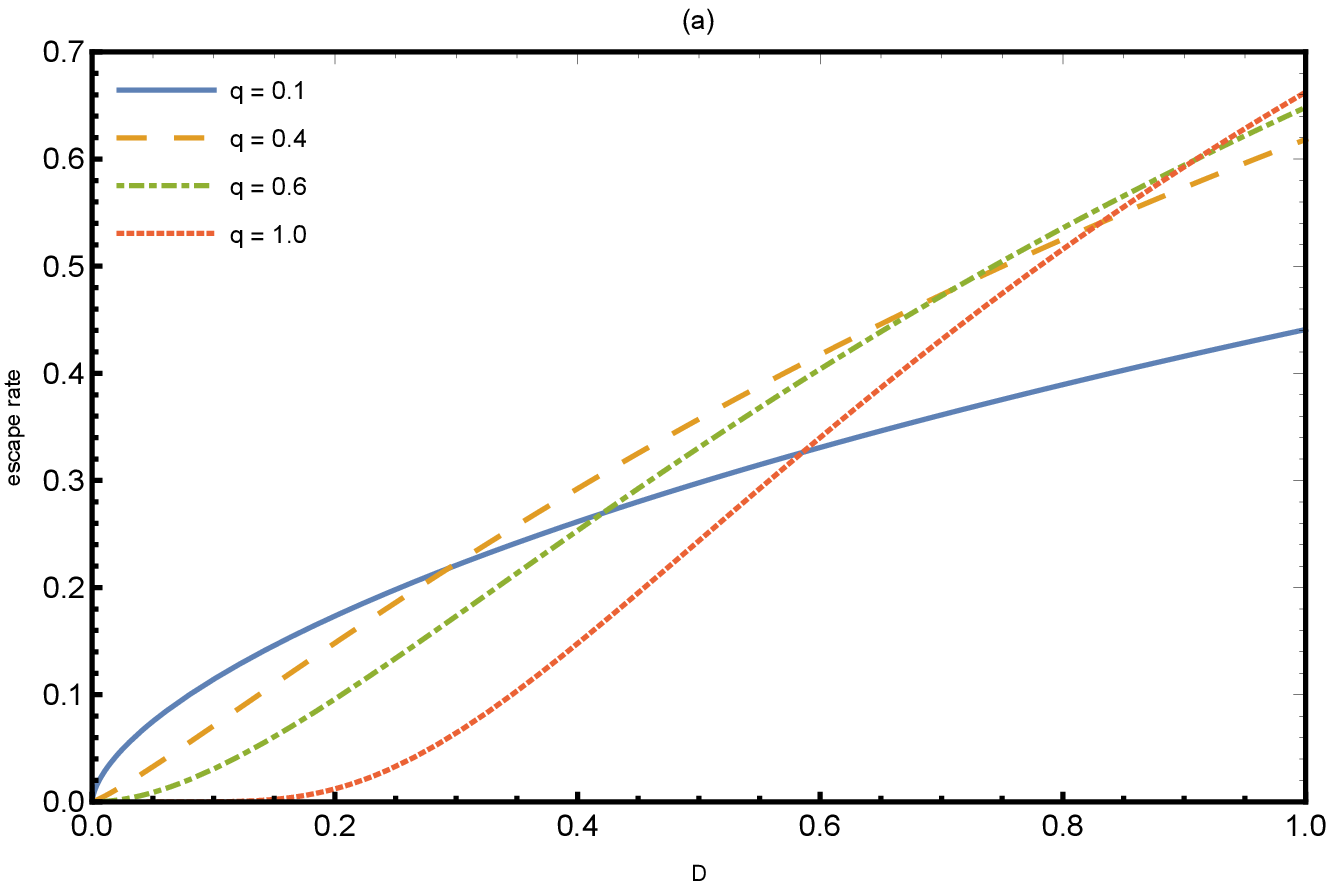}
\end{minipage}%
\begin{minipage}{0.5\textwidth}
\includegraphics[width=2.5in,height=2.2in]{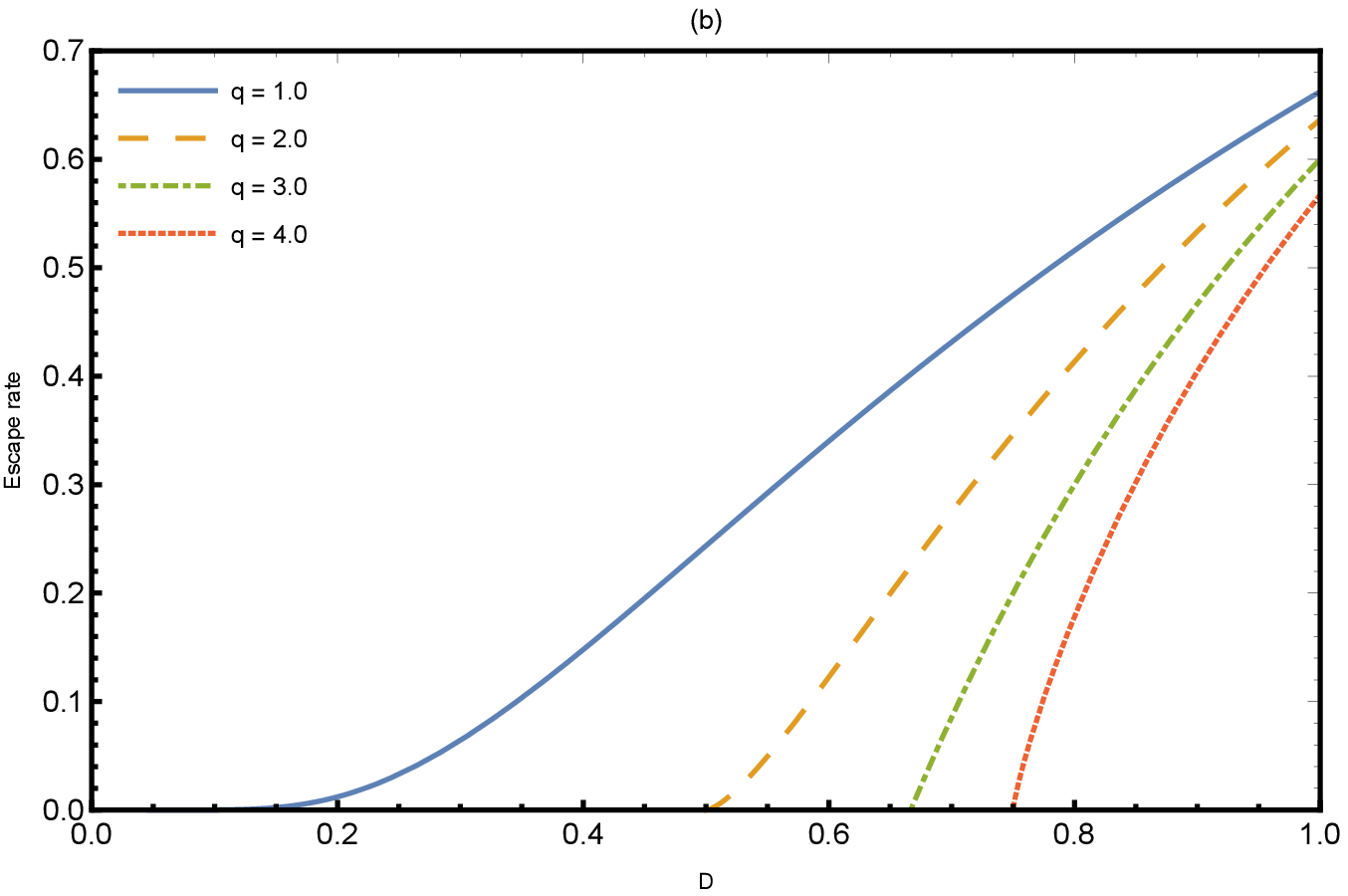}
\end{minipage}
 \caption{\label{f1}(Color online) The escape rate $r_{k}$ obtained from eq. (\ref{e8}) as a function of $D$ for various values of the the ratio $q=\nu/\mu >0$: (a) superdiffusive system and (b) subdiffusive system.} 
 \end{figure}

\begin{figure}\centering 
 \includegraphics[width=4.5in,height=2.8in]{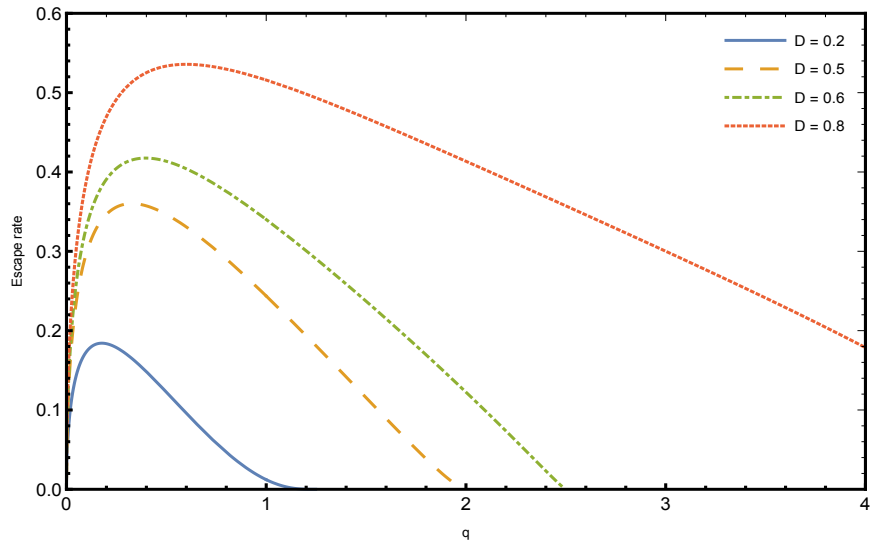}
 \caption{\label{f2}(Color online) The escape rate as a function of $q $ for some values of $D$.}
 \end{figure}

\begin{figure}\centering 
  \includegraphics[width=4.5in,height=2.8in]{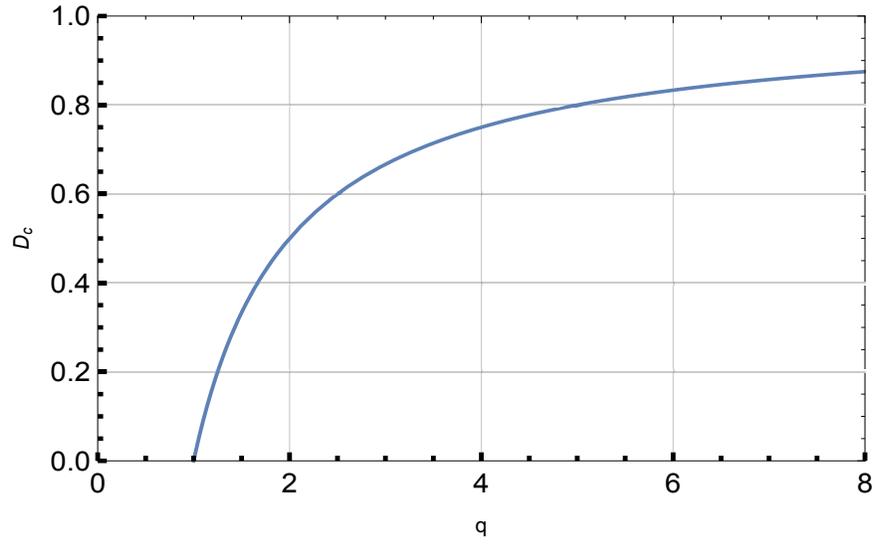}\label{f1c}
  \caption{(Color online) Critical noise strength $D_{c}$ as a function of $q$.} 
  \end{figure}

\begin{figure}\centering
\begin{minipage}{0.5\textwidth}
\includegraphics[width=2.5in,height=2.2in]{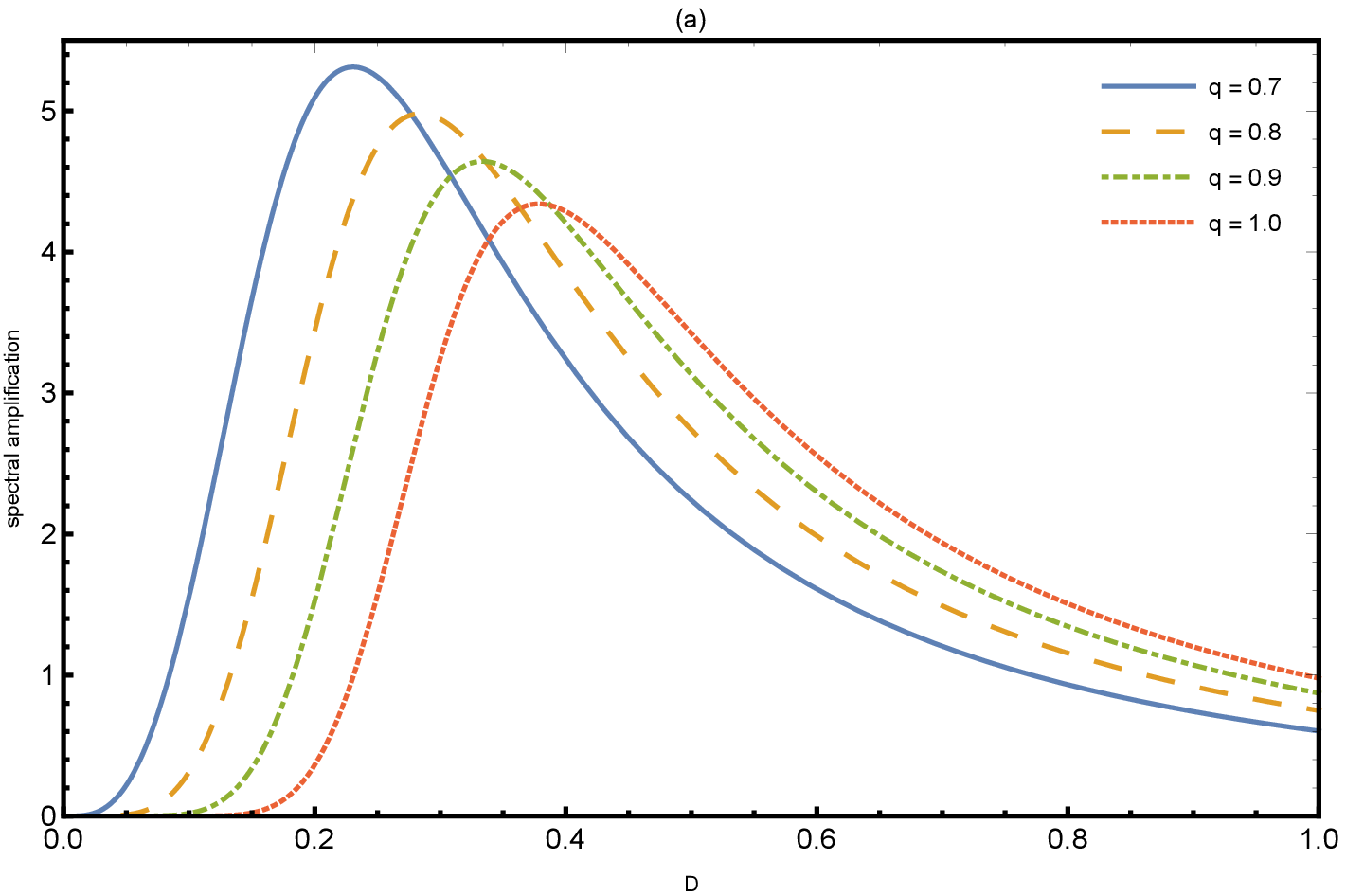}
\end{minipage}%
\begin{minipage}{0.5\textwidth}
\includegraphics[width=2.5in,height=2.2in]{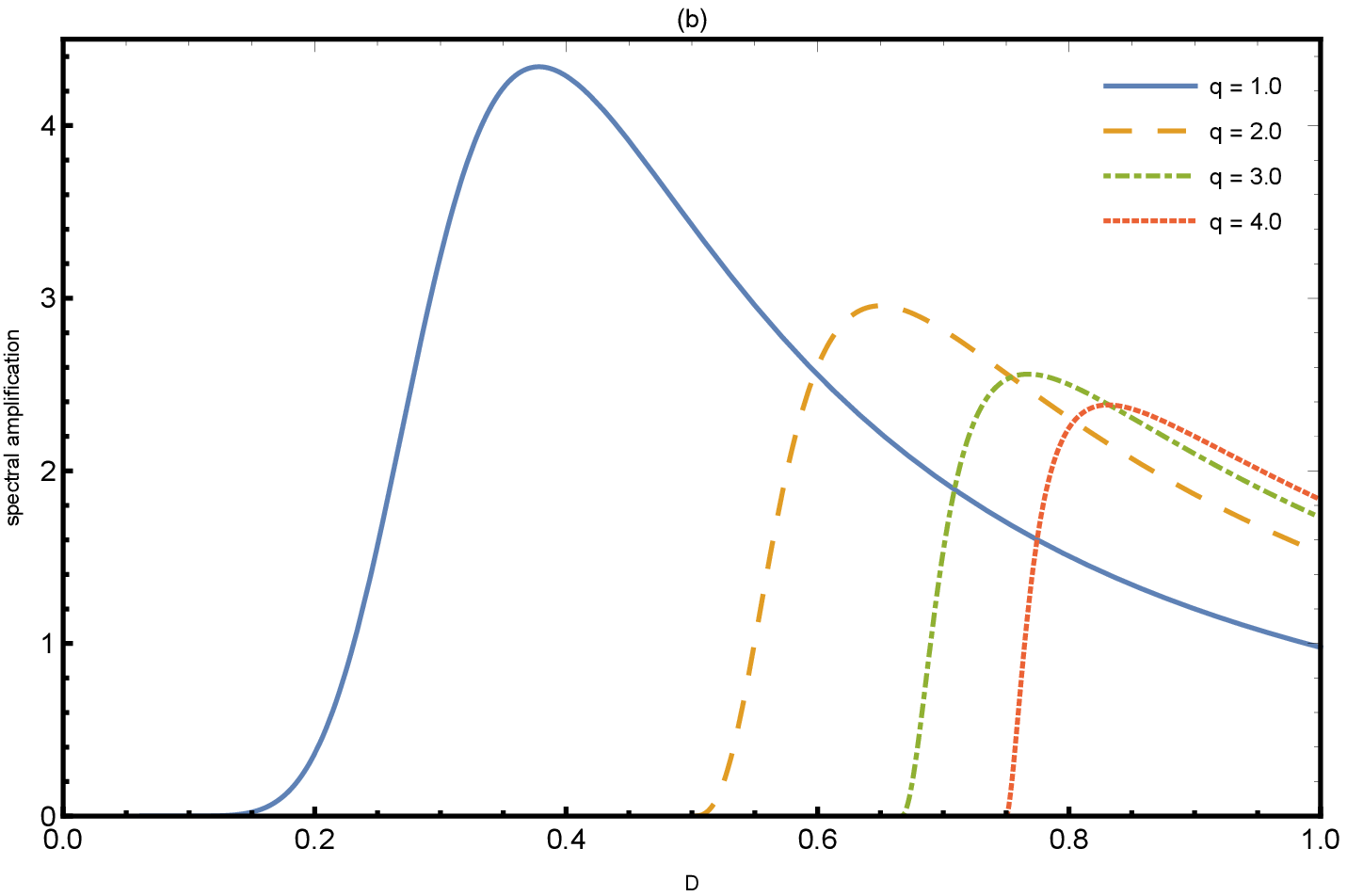}
\end{minipage}
\caption{\label{f3}(Color online) Spectral amplification versus $D$. (a) Superdiffusion, and (b) Subdiffusion.} 
\end{figure}

\begin{figure}\centering
\begin{minipage}{0.5\textwidth}
\includegraphics[width=2.5in,height=2.2in]{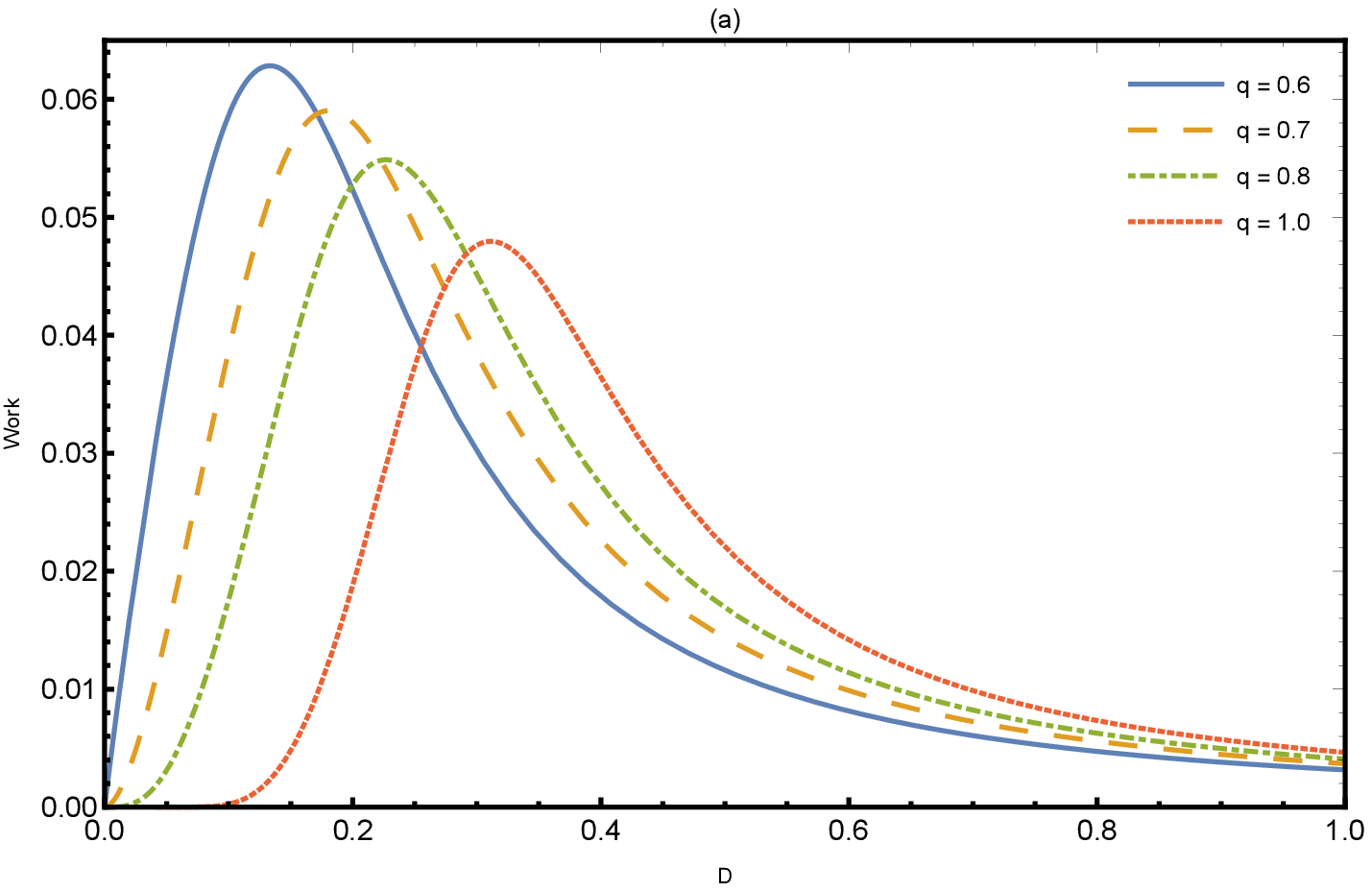}
\end{minipage}%
\begin{minipage}{0.5\textwidth}
\includegraphics[width=2.5in,height=2.2in]{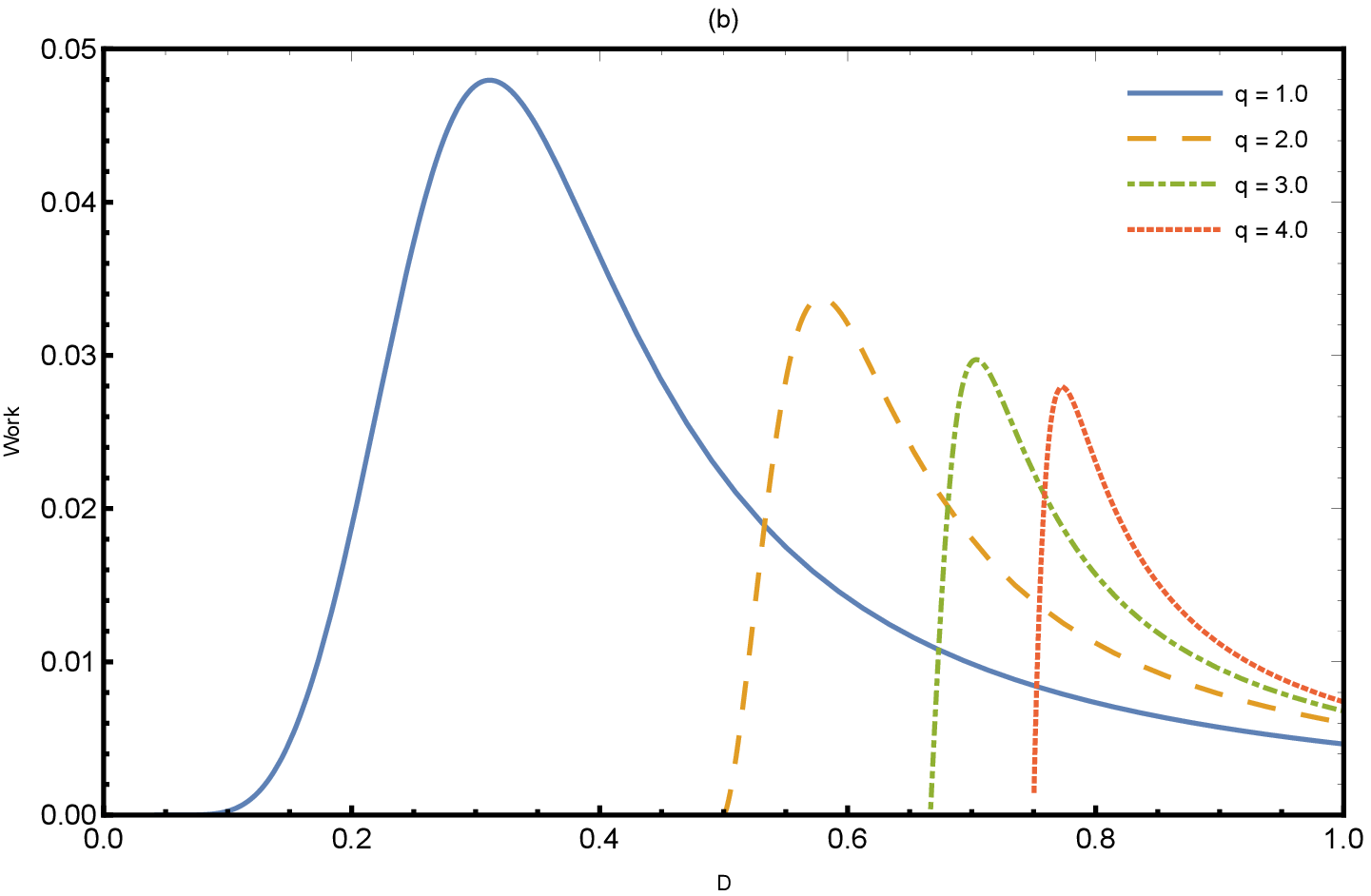}
\end{minipage}
\caption{\label{f4}(Color online) Average work versus $D$. (a) Superdiffusion, and (b) Subdiffusion.} 
\end{figure}

\begin{figure}\centering
\begin{minipage}{0.5\textwidth}
\includegraphics[width=2.5in,height=2.2in]{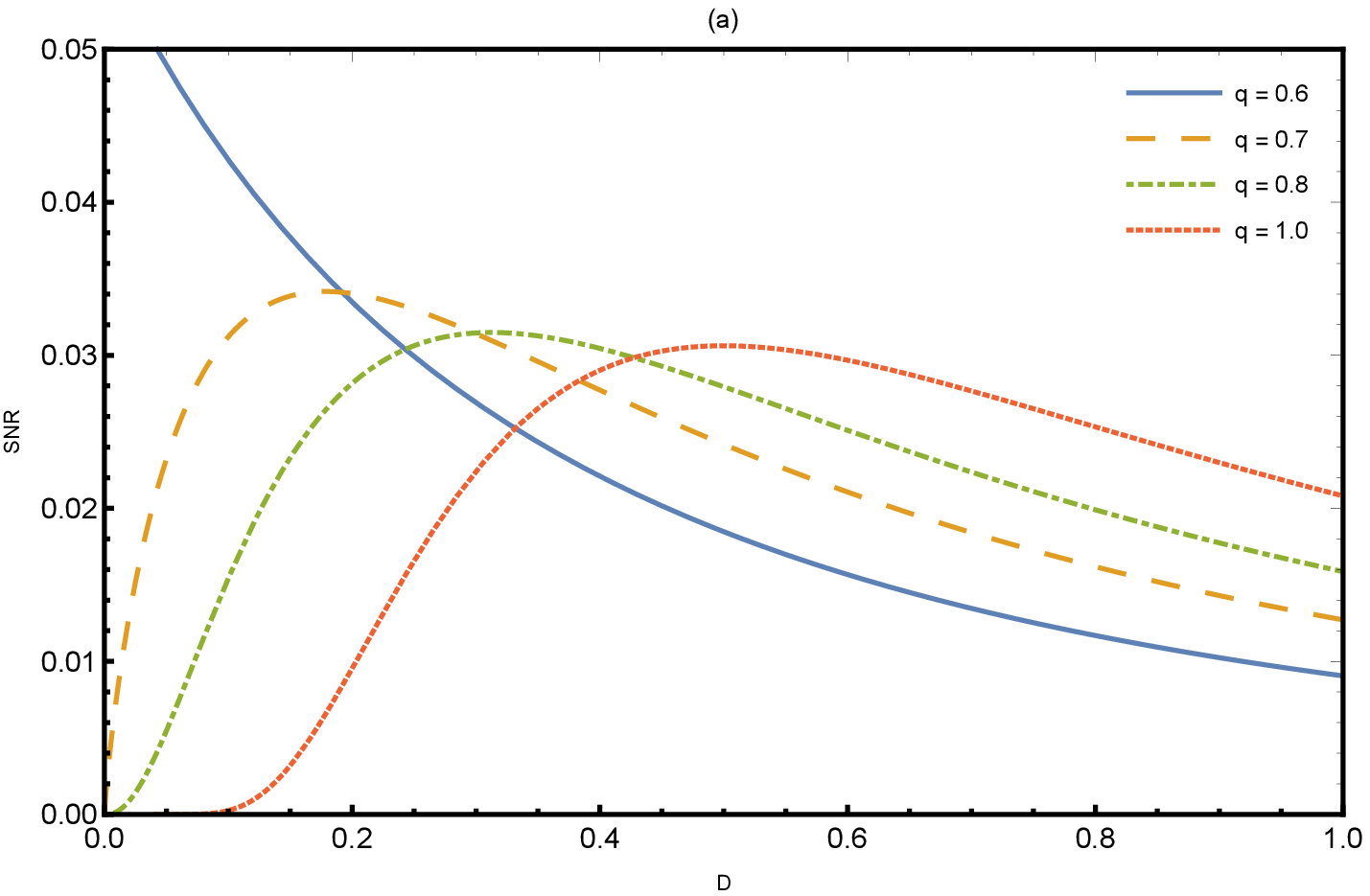}
\end{minipage}%
\begin{minipage}{0.5\textwidth}
\includegraphics[width=2.5in,height=2.2in]{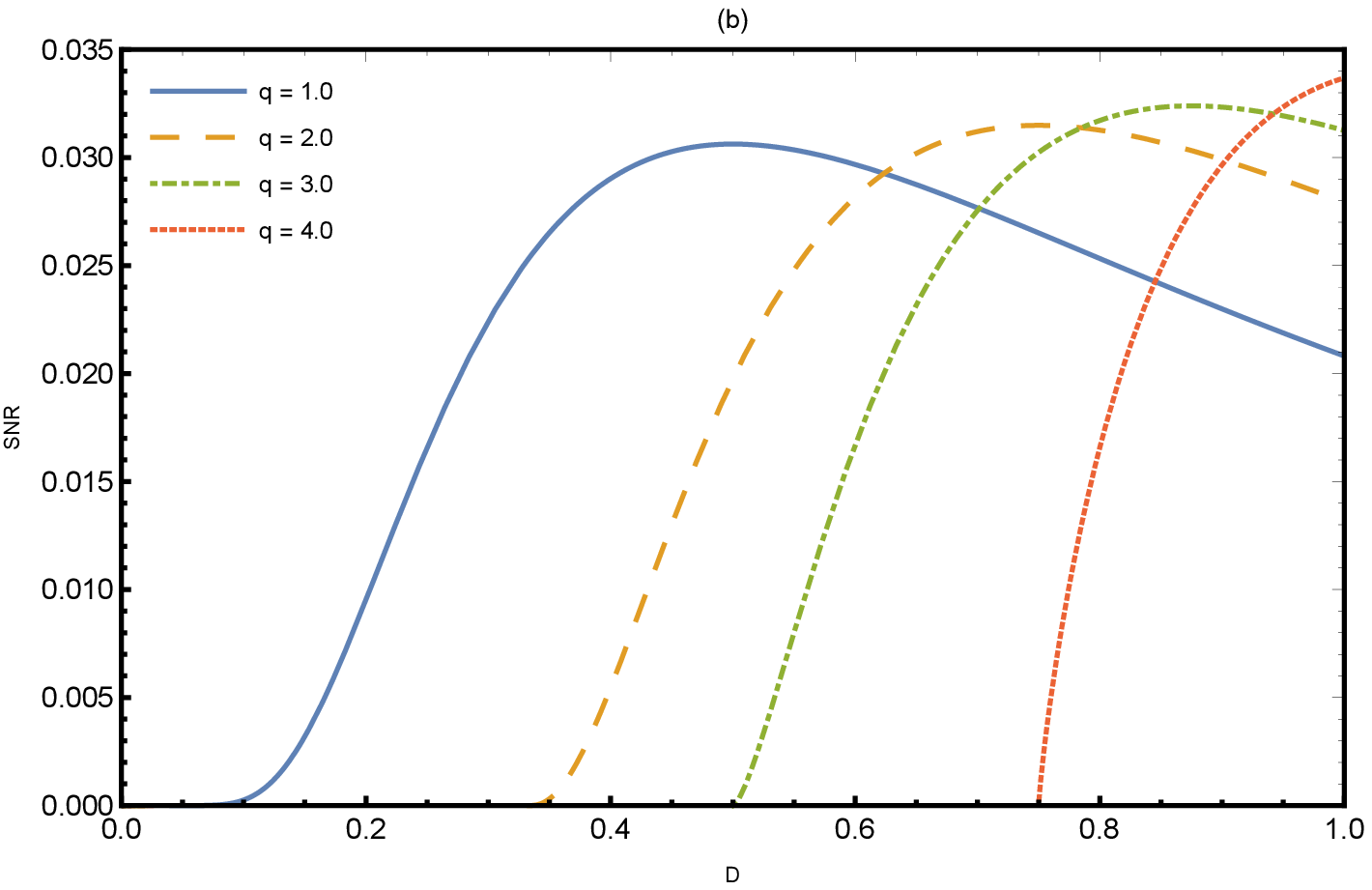}
\end{minipage}
\caption{\label{f5}(Color online) Signal-to-noise ratio versus $D$. (a) Superdiffusion, and (b) Subdiffusion .} 
\end{figure}

 In the previous section, we derived the expressions of relevant parameters characterizing SR in bistable systems driven by a periodic forcing, in the regime of anomalous diffusion. In the present section we shall examine their behaviors with variations of characteristic parameters of the model. \\
To start the escape rate $r_{k}$, given by eq. (\ref{e8}), is plotted in fig.\ref{f1} as a function of the noise strength $D$ in the two distinct diffusion regimes namely the superdiffusion regime (fig.1a) and the subdiffusion regime (fig.1b), for some values of the ratio $q=\nu/\mu$.  
 As one can notice, the escape rate increases with the noise strength following a power law for any value of the ratio $q$. However, for a fixed noise strength $D$, the escape rate will drop with $q$ in the subdiffusive regime while in the superdiffusive regime, there is a peak value in the escape rate for increasing $q$. The behavior of the escape rate with varying $q$ is more telling via the curves of fig. \ref{f2}, where $r_k$ is plotted over a broad range of values of $q$ for some values of $D$. 
 From figs. \ref{f1} and \ref{f2} it transpires that only in subdiffusive systems there exists a critical noise strength $D_{c}$, under which the particles will always be confined in one potential well. The existence of such critical noise strength can be derived from the relation $\beta_{\mu\nu}^{-1} = (\nu/\mu -1)E_b$, and finds its explanation in a cut-off of the stationary solution eq. (\ref{e3}) restricting the attainable space \cite{19}. The dependence of $D_{c}$ on parameters characterizing the diffusion anomaly i.e. $\nu$ and $\mu$, is illustrated in fig.\ref{f1c}.
 In fig.\ref{f3}, fig.\ref{f4} and fig.\ref{f5} respectively, we plotted the noise dependence of the spectral amplification, the average work and the signal-to-noise ratio, for some values of the ratio $q$. Here also we distinguished the superdiffusion regime from the subdiffusion regime in the graphical representations. 

\begin{figure}\centering 
\begin{minipage}{\textwidth}
\includegraphics[width=4.2in,height=2.8in]{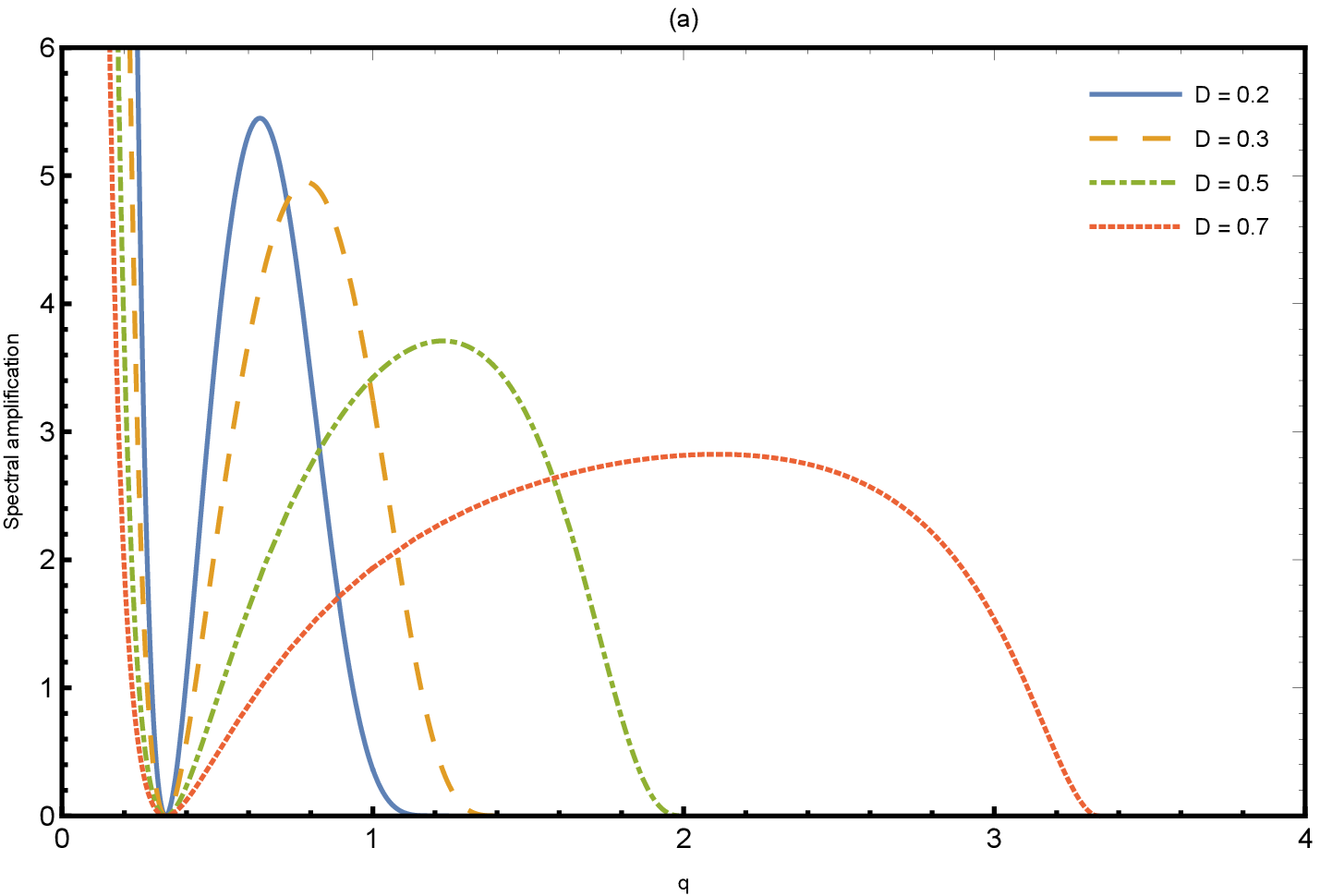}
\end{minipage}\\
\begin{minipage}{\textwidth}
\includegraphics[width=4.2in,height=2.8in]{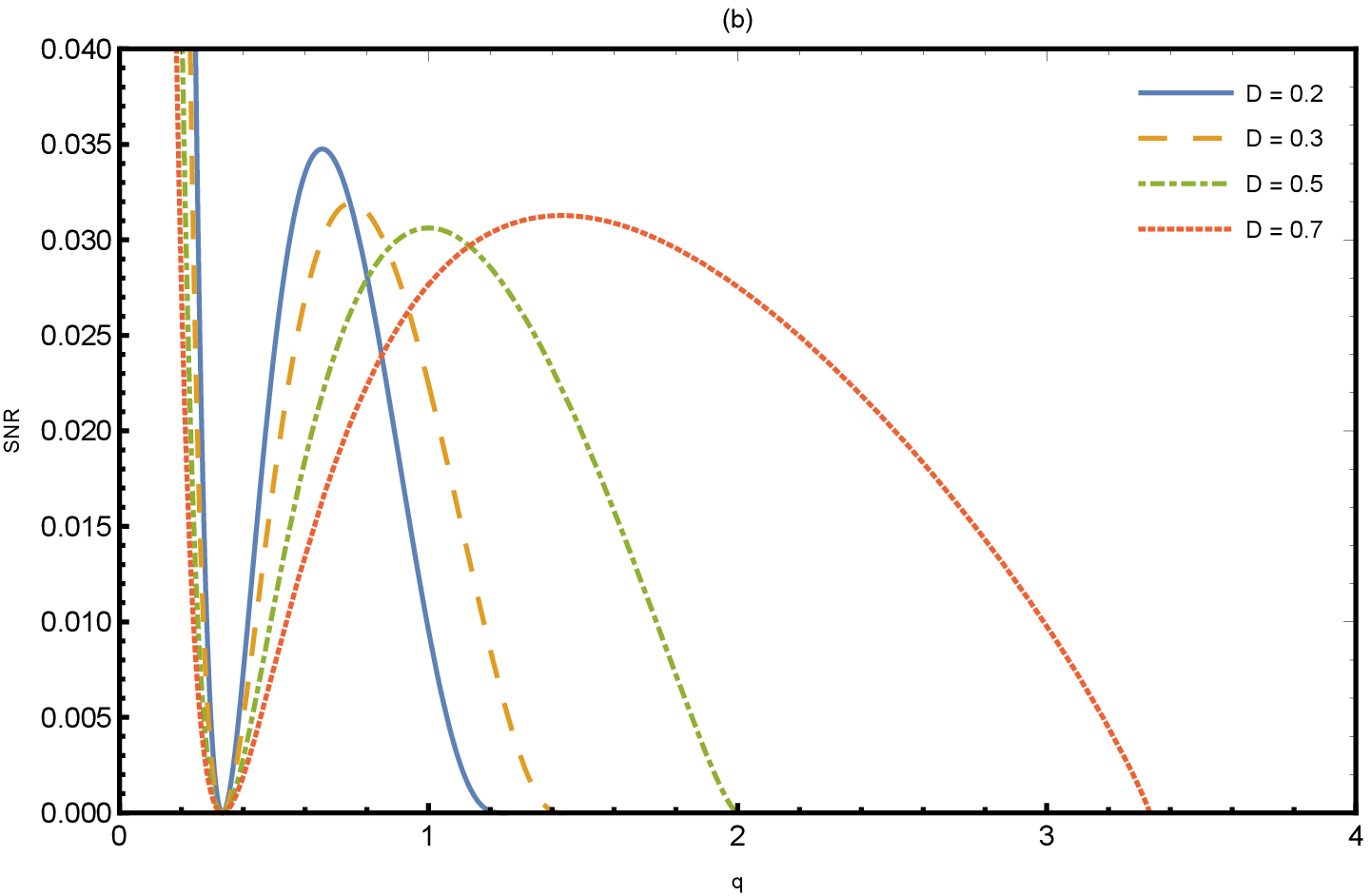}
\end{minipage}\\
\begin{minipage}{\textwidth}
\includegraphics[width=4.2in,height=2.8in]{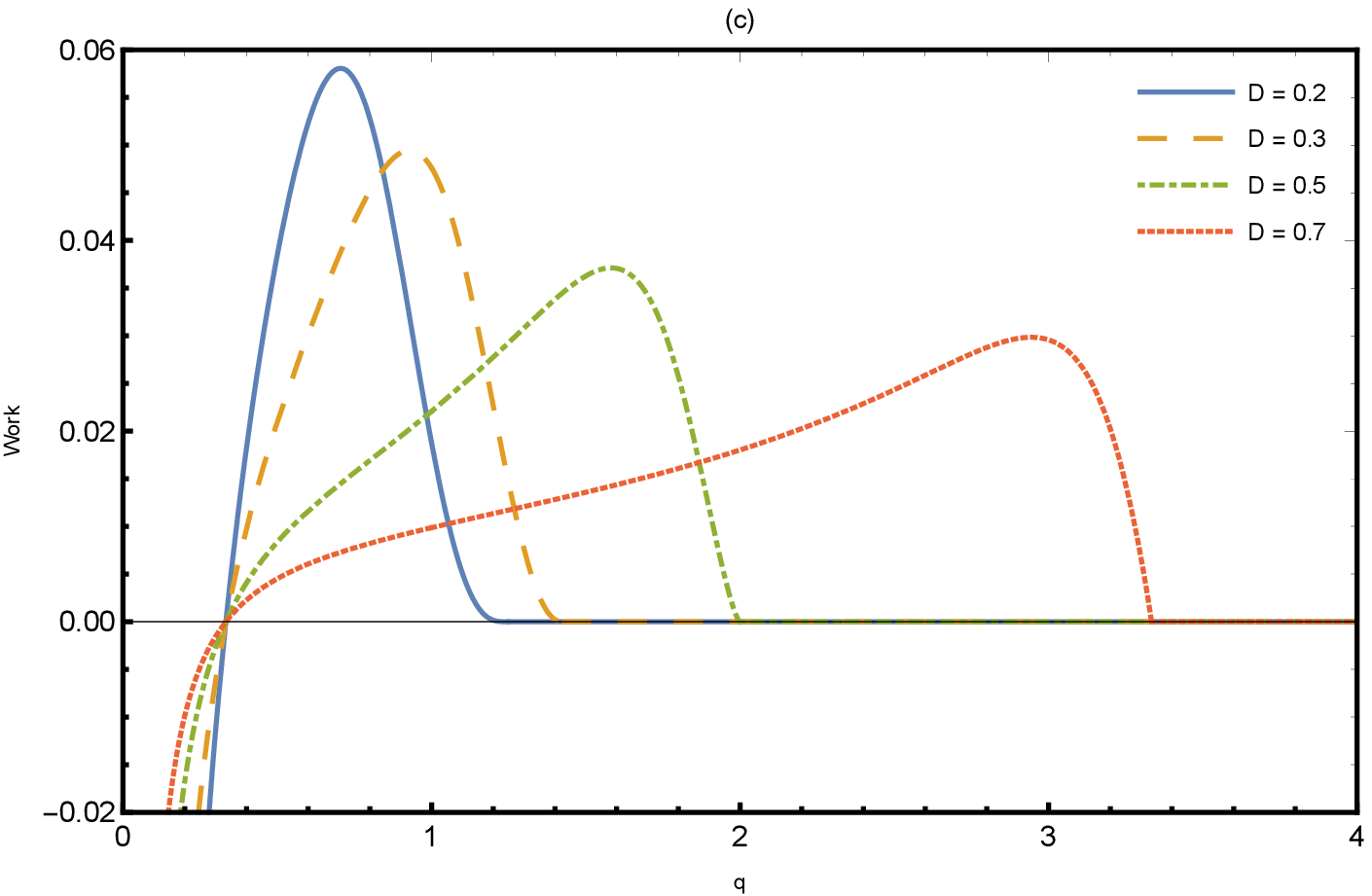}
\end{minipage}
\caption{\label{f6}(Color online) Plots of the quantifiers as function of $q$: (a) Spectral amplification, (b) Signal-to-noise ratio, and (c) average work.}
\end{figure}
The SPA and average work both present a resonant-like behavior for all values of $\nu$ and $\mu$, and qualitatively exhibit the same behaviors with variation of the noise strength. Indeed both show a decrease in their peak value and a rise of the resonance noise strength with the ratio $q$, for the selected range of values of $\nu$ and $\mu$, in either subdiffusive or superdiffusive regime. The cut-off noise is also reflected in the behavior of SPA and work in the superdiffusive systems, as it restricts the range of allowed noise strengths for the occurrence of SR. The behavior shown by the later quantifiers is also exhibited by the SNR, as shown in fig.\ref{f5} . However the peak value of the SNR reaches its minimum for $q=1 (\mu =\nu)$, and then rises as $q$ increases. It is also remarkable from curves in fig.\ref{f5}a and the analytical expression of the SNR that in the superdiffusive regime, there exists a range of values of $q$ setting the system in a configuration where SR can never occur. Similarly in fig.\ref{f5}b, there is a critical value of $q$ in the subdiffusion regime above which any stochastic resonance should not be considered to occur in the small-noise limit. In order to have a broader view on the impact of $\nu$ and $\mu$ on the three quantifiers, in fig. \ref{f6} we plotted the SPA, the SNR and the work as functions of $q$ considering some fixed values of $D$. From the figure we remark that at a fixed $D$ the quantifiers strongly diverge in the limit $q \rightarrow 0$, and are nullified for a specific configuration $q_{s}$. From their expressions in Eqs. (\ref{e17}), (\ref{e19}) and (\ref{e22}), this configuration is found to verify the relation $3\nu = \mu$. For all values of $D$ the quantifiers will all attain a peak value, then drastically drop to zero as $q$ increases beyond $q_{s}$. This drop can find its explanation in the fact that the increase of $q$ in the subdiffusive range set the system in a configuration where the noise strength can no more induce transitions between the degenerate stable states.

\section{conclusion}
We have studied the phenomenon of SR for bistable systems driven by a sinusoidal field, and subject to an anomalous diffusion represented by two real parameters i.e. $\nu$ and $\mu$. The model describes noise-driven transport phenomena involving either superdiffusive processes when $\mu > \nu$, normal diffusion when $\mu = \nu$ or subdiffusive processes when $\mu < \nu$ \cite{sok3,yina}. We predicted SR analytically via a set of quantifiers, including the spectral amplification, the average work done per cycle of the drive force and the signal-to-noise ratio. The occurrence of SR was found to be strongly affected by the order of anomaly of the system diffusion, furthermore there is a configuration of the system where the signal-to-noise ratio never shows SR for anomalous diffusion parameters in the range $3\mu < 5\nu$. It is important to point out that the peak positions of the three quantifiers were found to be crucially dependent on the order of diffusion anomaly, within the range of validity of the small-noise approximation. Our results establish for the first time, the relevant fact that taking into consideration the interactions of anomalous diffusive systems with a periodic signal, can provide a better understanding of the physics of stochastic resonance in bistable systems driven by periodic forces. \\
A possible extension of the present study would be to look at the effects of deformability of the bistable potential, such as the change of confinement of the potential well (as done in ref. \cite{ram1}), or of positions of the double-degenerate potential minima \cite{dik1}, on profiles of the stochastic-resonance quantifiers. Indeed the bistable potential $U(x)$ considered in the present work is the so-called $\phi^4$ \cite{schrif,dik,dika} whose rigid profile, reflected in its fixed minima positions and the fixed barrier height, limit their applicability to systems with soft profiles as it is common in polymers and biophysical systems. An analysis of conditions for the occurrence of stochastic resonance in bistable systems with deformable double-well shapes \cite{dik1}, should provide novel insight onto the physics of systems of this specific class from a general standpoint.

\end{document}